\documentclass[a4paper,10pt,3p,onecolumn,compress]{elsarticle}
\usepackage[T1]{fontenc}
\usepackage{graphicx} 
\usepackage{amsmath}
\usepackage{amssymb}
\usepackage{xspace}
\usepackage{orcidlink}
\usepackage{lineno}
\usepackage{booktabs}
\usepackage{multirow}
\usepackage{multicol}

\title{Precision cross-sections for advancing cosmic-ray physics. \\
Input to the 2026 ESPPU from the XSCRC community}
\date{\today}

\newcommand{\sqrtsnn}{\sqrt{s_\text{NN}}}
\newcommand{\pt}{\ensuremath{p_{\mathrm{T}}}\xspace}

\newcommand{\neutron}{\ensuremath{\mathrm{n}}\xspace}
\newcommand{\proton}{\ensuremath{\mathrm{p}}\xspace}

\newcommand{\deuteron}{\ensuremath{\mathrm{d}}\xspace}
\newcommand{\positron}{\ensuremath{\mathrm{e^+}}\xspace}
\newcommand{\helium}{\ensuremath{\mathrm{He}}\xspace}
\newcommand{\antiproton}{\ensuremath{\mathrm{\overline{p}}}\xspace}
\newcommand{\antineutron}{\ensuremath{\mathrm{\overline{n}}}\xspace}
\newcommand{\antideuteron}{\ensuremath{\mathrm{\overline{d}}}\xspace}
\newcommand{\antitriton}{\ensuremath{\mathrm{\overline{t}}}\xspace}
\newcommand{\antihelium}{\ensuremath{\mathrm{\overline{He}}}\xspace}
\newcommand{\antiheliumthree}{\ensuremath{\mathrm{^3\overline{He}}}\xspace}
\newcommand{\antiheliumfour}{\ensuremath{\mathrm{^4\overline{He}}}\xspace}

\newcommand{\Lbar}{\ensuremath{\mathrm{\overline{\Lambda}}}\xspace}
\newcommand{\Sbar}{\ensuremath{\mathrm{\overline{\Sigma}}}\xspace}

\begin{document}

\author[1]{S.~Mariani\corref{cor1}\orcidlink{0000-0002-7298-3101}}%
\ead{saverio.mariani@cern.ch}
\cortext[cor1]{Corresponding author}
\author[2]{L.~Audouin\orcidlink{0000-0001-9899-6923}}
\author[3]{E.~Berti\orcidlink{0000-0002-5841-7760}}
\author[4]{P.~Coppin\orcidlink{0000-0001-6869-1280}}
\author[5]
{M.~Di Mauro\orcidlink{0000-0003-2759-5625}}
\author[6]{P.~von~Doetinchem\orcidlink{0000-0002-7801-3376}}
\author[1,5,7]{F.~Donato\orcidlink{0000-0002-3754-3960}}
\ead{fiorenza.donato@unito.it}
\author[8,9]{C.~Evoli\orcidlink{0000-0002-6023-5253}}
\author[10]{Y.~Génolini\orcidlink{0000-0002-7326-1282}}
\author[11]{P.~Ghosh\orcidlink{0000-0003-1293-3660}}
\author[12]{I.~Leya\orcidlink{0000-0002-3843-6681}}
\author[13,14]{M.~J.~Losekamm\orcidlink{0000-0001-7854-2334}}
\author[15]{D.~Maurin\orcidlink{0000-0002-5331-0606}}
\ead{dmaurin@lpsc.in2p3.fr}
\author[16]{J.~W.~Norbury\orcidlink{0009-0000-0740-0611}}
\author[17,18]{L.~Orusa\orcidlink{0000-0002-1879-457X}}
\author[4]{M.~Paniccia\orcidlink{0000-0001-8482-2703}}
\author[1]{T.~Poeschl\orcidlink{0000-0003-3754-7221}}
\author[10]{P.~D.~Serpico\orcidlink{0000-0002-8656-7942}}
\author[4]{A.~Tykhonov\orcidlink{0000-0003-2908-7915}}
\author[19]{M.~Unger\orcidlink{0000-0002-7651-0272}}
\author[20]{M.~Vanstalle\orcidlink{0000-0003-1193-7475}}
\author[21,22]{M.-J.~Zhao\orcidlink{0000-0001-6844-1409}}
\author[23,9]{D.~Boncioli\orcidlink{0000-0003-1186-9353}}
\author[5,7]{M.~Chiosso\orcidlink{0000-0001-6994-8551}}
\author[5]{D.~Giordano\orcidlink{0000-0003-0228-9226}}
\author[24]{D.~M.~Gomez~Coral\orcidlink{0000-0002-9200-6607}}
\author[3]{G.~Graziani\orcidlink{0000-0001-8212-846X}}
\author[1]{C.~Lucarelli\orcidlink{0000-0002-8196-1828}}
\author[25,26]{P.~Maestro\orcidlink{0000-0002-4193-1288}}
\author[13]{M.~Mahlein\orcidlink{0000-0003-4016-3982}}
\author[27]{L.~Morejon\orcidlink{0000-0003-1494-2624}}
\author[28]{J.~Ocampo-Peleteiro\orcidlink{0000-0002-7585-2641}}
\author[28]{A.~Oliva\orcidlink{0000-0002-6612-6170}}
\author[20]{T.~Pierog\orcidlink{0000-0002-7472-8710}}
\author[1]{L.~Šerkšnytė\orcidlink{0000-0002-5657-5351}}

\affiliation[1]{organization={European Organization for Nuclear Research (CERN)}, city={Geneva}, country={Switzerland}}
\affiliation[2]{organization={Université Paris-Saclay, IJC Lab, CNRS/IN2P3}, postcode={91405}, city={Orsay}, country={France}}
\affiliation[3]{organization={INFN, Sezione di Firenze, I-50019 Sesto Fiorentino, Florence, Italy}}
\affiliation[4]{organization={DPNC, Université de Genève}, postcode={1211 Genève 4}, city={Geneva}, country={Switzerland}}
\affiliation[5]{INFN, Sezione di Torino, Via P. Giuria 1, 10125 Torino, Italy}
\affiliation[6]{organization={University of Hawaii at Manoa}, postcode={HI}, city={Honolulu}, country={USA}}
\affiliation[7]{organization={Università degli Studi di Torino}, city={Torino}, country={Italy}}
\affiliation[8]{organization={Gran Sasso Science Institute (GSSI)}, addressline={Viale Francesco Crispi 7}, postcode={67100}, city={L’Aquila}, country={Italy}}
\affiliation[9]{organization={INFN-Laboratori Nazionali del Gran Sasso (LNGS)}, addressline={via G. Acitelli 22}, postcode={67100}, city={Assergi (AQ)}, country={Italy}}
\affiliation[10]{organization={LAPTh, CNRS, Université Savoie Mont Blanc}, postcode={F-74940}, city={Annecy}, country={France}}
\affiliation[11]{organization={NASA Goddard Space Flight Center, Greenbelt, Maryland, 20771, USA}} 
\affiliation[12]{organization={University of Bern, Space Sciences and Planetology}, postcode={CH-3012}, city={Bern}, country={Switzerland}}
\affiliation[13]{organization={Technical University of Munich, School of Natural Sciences}, city={Garching}, country={Germany}}
\affiliation[14]{organization={Excellence Cluster ORIGINS}, city={Garching}, country={Germany}}
\affiliation[15]{organization={LPSC, Université Grenoble-Alpes, CNRS/IN2P3}, postcode={38026}, city={Grenoble}, country={France}}
\affiliation[16]{organization={NASA Langley Research Center, Hampton, Virginia, 23666, USA}}
\affiliation[17]{organization={Department of Astrophysical Sciences, Princeton University}, postcode={NJ 08544}, city={Princeton}, country={USA}}
\affiliation[18]{organization={Department of Physics and Columbia Astrophysics Laboratory, Columbia University}, postcode={NY 10027}, city={New York}, country={USA}}
\affiliation[19]{organization={IAP, KIT}, city={Karlsruhe}, country={Germany}}
\affiliation[20]{organization={Université de Strasbourg, CNRS, IPHC-UMR7178, F-67000 Strasbourg, France}}
\affiliation[21]{organization={Key Laboratory of Particle Astrophysics, Institute of High Energy Physics, Chinese Academy of Sciences}, postcode={100049}, city={Beijing}, country={China}}
\affiliation[22]{organization={China Center of Advanced Science and Technology, 100190, Beijing, China}}
\affiliation[23]{organization={Università degli Studi dell’Aquila, Dipartimento di Scienze Fisiche e Chimiche, Via Vetoio, 67100, L’Aquila, Italy}}
\affiliation[24]{organization={Instituto de F{\'i}sica, Universidad Nacional Autónoma de México, Circuito de la Investigación Cient{\'i}fica, Ciudad de México, México}}
\affiliation[25]{organization={Department
of Physical Sciences, Earth and Environment, University of Siena, via Roma 56, 53100
Siena, Italy}}
\affiliation[26]{organization={INFN, Sezione di Pisa, Polo Fibonacci, Largo B. Pontecorvo 3, 56127 Pisa, Italy}}
\affiliation[27]{organization={Bergische Universität Wuppertal,
Gausstrasse 20, 42117 Wuppertal, Germany}}
\affiliation[28]{organization={INFN, Sezione di Bologna, 40126 Bologna, Italy}}

\begin{abstract}
The latest generation of cosmic-ray direct detection experiments is providing a wealth of high-precision data, stimulating a very rich and active debate in the community on the related strong discovery and constraining potentials on many topics, namely dark matter nature, and the sources, acceleration, and transport of Galactic cosmic rays. However, interpretation of these data is strongly limited by the uncertainties on nuclear and hadronic cross-sections. This contribution is one of the outcomes of the \textit{Cross-Section for Cosmic Rays at CERN} workshop series, that built synergies between experimentalists and theoreticians from the astroparticle, particle physics, and nuclear physics communities. A few successful and illustrative examples of CERN experiments' efforts to provide missing measurements on cross-sections are presented. In the context of growing cross-section needs from ongoing, but also planned, cosmic-ray experiments, a road map for the future is highlighted, including overlapping or complementary cross-section needs from applied topics (e.g., space radiation protection and hadrontherapy).
\end{abstract}
\maketitle

\section{Introduction}
Charged particles arriving at the Earth from space with energies above 100 MeV -- the so-called cosmic rays (CRs) -- have always represented a complementary facility to accelerators to advance the understanding of the Standard Model~(SM) of Particle Physics. The latest generation of direct-detection space-based experiments (AMS~\cite{AMS:PhysRep2021}, CALET~\cite{CALET}, DAMPE~\cite{DampeMission}, ISS-CREAM~\cite{ISS-CREAM:Choi2022}, and PAMELA~\cite{2017NCimR..40..473P}) have marked the entrance to a high-precision era for CR physics, reaching the few percent level precision from GeV to hundreds of TeV energies. The related discovery potential for the astroparticle field, but also to address fundamental questions still unanswered by the SM -- among all the nature of Dark Matter~(DM) --, is undoubted. Currently, the interpretation of these data is unfortunately limited by the precision at which production and nuclear fragmentation cross-sections~(XS) are known, i.e., in the 10 to 20\% precision at most. High-Energy Physics (HEP) experiments at accelerators, especially at the CERN complex, have started to provide such measurements, in some cases even going beyond their original physics goals, building a very successful synergy with theoreticians and experimentalists of the astroparticle community. A clear example is the XSCRC (Cross-Section for Cosmic Rays at CERN) workshop series, whose third edition took place in October 2024. During this workshop, an actionable programme for the next decade was discussed, which lead to the writing of a detailed and comprehensive road map~\cite{XSCRC24_paper}.

In this contribution to the 2026 European Strategy for Particle Physics Update~(ESPPU), an excerpt of the above road map is presented. Because of the length constraint, several societal topics presented in Ref.~\cite{XSCRC24_paper} are skipped -- namely cosmogenic production in meteorites related to life on Earth \cite{DavidLeya2019}, space exploration and radiation protection \cite{Norbury2012}, and hadrontherapy \cite{Battistoni2021} --, but are nonetheless of relevance for the ESPPU.
This contribution focuses on the needs for DM detection with Galactic CRs (GCRs), and is organised as follows. In Sec.~\ref{sec:motivation}, some of the physics cases opened by the high-precision CR data are described, and the crucial missing inputs to address them are listed. In Sec.~\ref{sec:synergies}, successful examples of CERN experiments results, pertaining to these physics cases, are given, together with their future prospects. Finally, in Sec.~\ref{sec:conclusions}, a list of recommendations to further strengthen the presented physics programme is discussed.

\section{Physics cases in a cosmic-ray precision era}
\label{sec:motivation}
\begin{figure}
	\centering
    \includegraphics[width=0.48\textwidth]{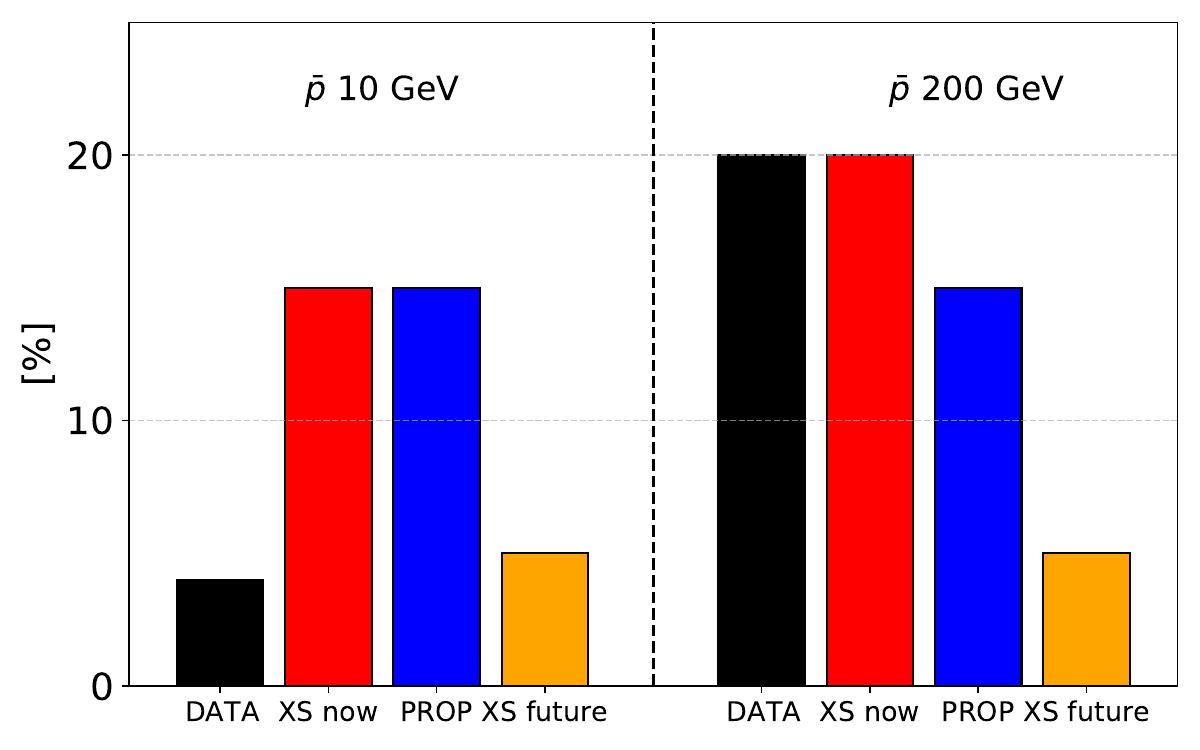}
	\includegraphics[width=0.48\textwidth]{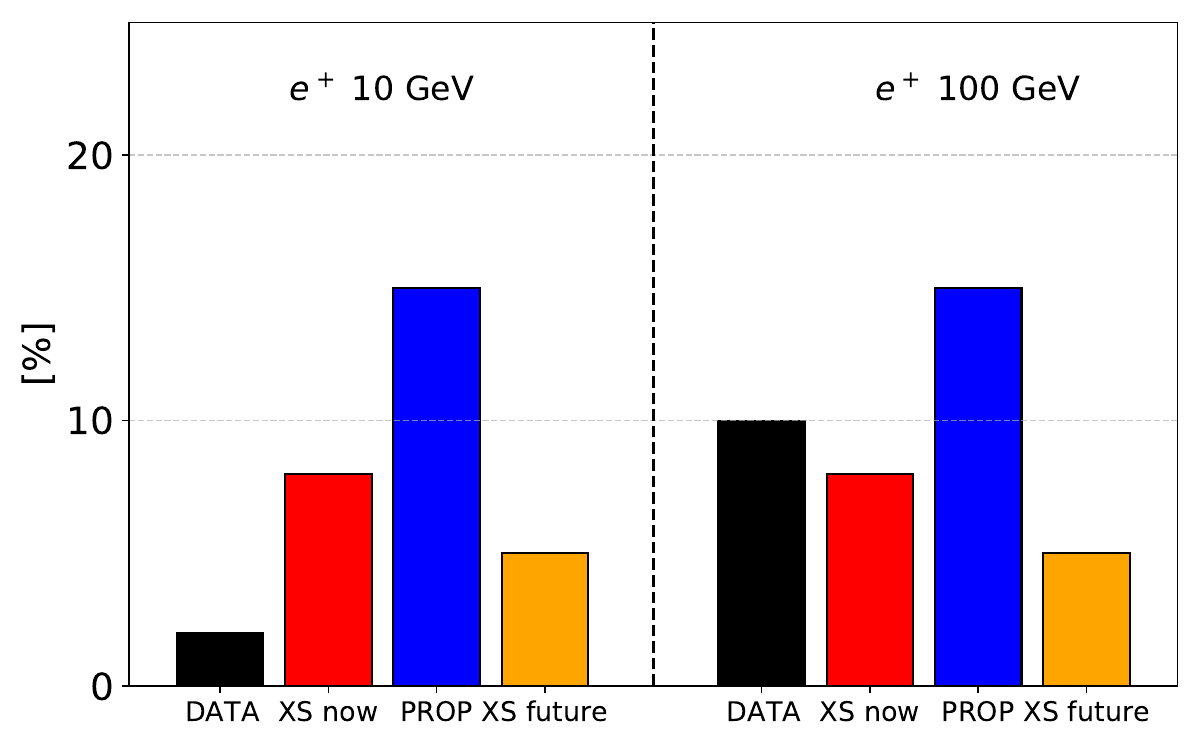}
	\caption{\label{fig:XS_cases}Illustration, on CR \antiproton (left) and \positron (right), of current data precision (black), compared to current XS (red) and propagation (blue) precision, and desired XS precision (orange).}
\end{figure}

The so-called indirect search for a DM particle aims at uncovering excesses of the antimatter particle fluxes in CRs with respect to the modelled astrophysical background. This {\em secondary} background originates from interactions of {\em primary} GCRs (i.e., present in sources), mainly \proton and \helium nuclei, with the interstellar medium~(ISM) made of H (90\%) and He (10\%). Figure~\ref{fig:XS_cases} illustrates, on \antiproton and \positron, that new high-precision nuclear data are needed to get the modelled astrophysical fluxes on par with CR data precision. A summary of the needed XS, to precisely constrain the secondary antimatter fluxes and to discriminate a possible DM contribution, is reported in Table~\ref{tab:measurements}, with the main cases detailed below.

\begin{table}
    \setlength{\tabcolsep}{6pt}
    \centering
    \caption{Summary of the wish list of XS, for CRs that can be indirect probes of particle DM.
    Here, $n_{\rm tot}$ is the integrated multiplicity. The most pressing need is for \antiproton, whose interpretation is already limited by XS uncertainties, but forthcoming data for \antideuteron and possible \antihelium events from AMS call for new XS measurements as well.\vspace{1.5mm}}
    \label{tab:measurements}
    \centering
    \begin{tabular}{clcccc}
        \toprule
        \textbf{Particle} & \textbf{Reaction} & \textbf{Measurement} & \textbf{$\sqrt{s}$} & \textbf{Sought precision}\\
        \midrule
        \multirow{6}{*}{\antiproton} & $\proton+\proton\rightarrow\antiproton+{\rm X}$ & \multirow{6}{*}{$\sigma_{\rm inv}$} & \multirow{6}{*}{5 to 100 GeV} & $<3\%$\\
         & $\proton+\helium\rightarrow\antiproton+{\rm X}$ & & & $<5\%$\\
         & $\proton+\proton\rightarrow\overline{\Lambda}+{\rm X}$ & & & $<10\%$\\
         & $\proton+\helium\rightarrow\overline{\Lambda}+{\rm X}$ & & & $<10\%$\\
         & $\proton+\proton\rightarrow \antineutron+{\rm X}$ & & & $<5\%$ \\
         & $\proton+\neutron\rightarrow\antiproton+{\rm X}$ & & & $<5\%$\\
        \midrule
        \multirow{2}{*}{\antideuteron} & $\proton+\proton\rightarrow\antideuteron+{\rm X}$ & $\sigma_{\rm inv}/n_{\rm tot}$ & 5 to 100 GeV & (any data)\\
        & $\proton+\helium\rightarrow\antideuteron+{\rm X}$ & $\sigma_{\rm inv}/n_{\rm tot}$ & 5 to 100 GeV & (any data)\\
         & $\antiproton+\proton\rightarrow\antideuteron+{\rm X}$ & $\sigma_{\rm inv}$ & 2 to 10 GeV & (any data) \\
        \midrule
        \multirow{1}{*}{\antihelium} & $\proton+\proton\rightarrow\antihelium+{\rm X}$ & $\sigma_{\rm inv}/n_{\rm tot}$ & 5 to 100 GeV & (any data)\\
        \midrule
        \multirow{2}{*}{$e^\pm$} & $\proton+\helium\rightarrow\pi^\pm+{\rm X}$ & \multirow{2}{*}{$\sigma_{\rm inv}$} & \multirow{2}{*}{5 to 100 GeV} & $<5\%$\\
        & $\proton+\helium\rightarrow K^\pm+{\rm X}$ & & & $<5\%$ \\
        \midrule
        \multirow{2}{*}{$\gamma$} & $\proton+\proton\rightarrow\pi^0+{\rm X}$ & \multirow{2}{*}{$\sigma_{\rm inv}$} & \multirow{2}{*}{5 to 1000 GeV} & $<5\%$\\
        & $\proton+\helium\rightarrow\pi^0+{\rm X}$ & & & $<5\%$\\
        \bottomrule
    \end{tabular}
\end{table}

\paragraph{Antiprotons} \antiproton are mainly expected to be of secondary origin, as no primary source is known. A few percent precision measurement of their flux, up to 500\,GeV energy values, has been published based on 6.5~years of data collected by the AMS experiment \cite{AMS:PhysRep2021}. Based on the current predictions, AMS data are globally consistent with the secondary-only production hypothesis \cite{Boudaud:2019efq}, but a possible excess around 10\,GeV cannot be further tested because of modelling uncertainties dominated by XS. Reducing them by exploiting collisions, on both H and He targets, is therefore of pivotal importance. The \antiproton production XS can be expressed as the sum of a prompt emission, a delayed contribution due to decays of the \Lbar and \Sbar anti-hyperons, and one coming from the decay of antineutrons, \antineutron~\cite{Winkler:2017xor}:
        \begin{equation}
         \sigma_{\rm inv}^{\antiproton} = \sigma_{\rm inv}^{\antiproton,\antineutron\;\rm prompt} + \sigma_{\rm inv}^{\antiproton,\antineutron\;\rm delayed} =\sigma_{\rm inv}^{\antiproton\;\rm prompt}  (2+\Delta_{\rm IS}+ 2 \Delta_{\Lambda})\,.
        \label{eq:def_pbarinv}
\end{equation}
For the prompt and delayed contributions, a few measurements exist, but they should be extended towards lower collision energy and to precisions below 3\% and 5\%, respectively. The main model uncertainty currently comes from the poor knowledge of $\Delta_{\rm IS}$. The NA49 experiment reported indeed an asymmetry up to 50\% in the antineutron to antiproton production~\cite{Fischer:2003xh}, although affected by large uncertainties. An independent measurement for this quantity, with uncertainties below 5\% and in the nucleon-nucleon centre-of-mass energy range $\sqrtsnn= 5\text{--}110$\,GeV, would therefore be highly desirable;

\paragraph{Antinuclei} While not yet observed in GCRs, the case for light anti-nuclei, namely \antideuteron and \antihelium, is even more promising, given that their production in GCR-ISM interactions below the GeV energy is suppressed by one order of magnitude with respect to the yields associated to viable thermal relic DM models. The main theoretical uncertainty is related to the description of nucleons coalescence to form light anti-nuclei, only constrained by scarce measurements. In the context of the awaited AMS and GAPS~\cite{2023APh...14502791R} results, new nuclear data with uncertainties $\lesssim 30\%$ at $\sqrtsnn < 100$\,GeV energy would strictly constrain the expected \antideuteron and \antihelium secondary flux. In addition, measurements of the inelastic XS affecting the \antideuteron and \antihelium fluxes during propagation are also desired.

\paragraph{Positron and $\gamma$-ray probes}
The rising \positron fraction above 10\,GeV, as measured by PAMELA~\cite{PAMELA:2008gwm} and AMS~\cite{PhysRevLett.113.121101}, raised a vivid debate concerning the origin of the primary sources of these high-energy \positron (pulsar wind nebulae, DM annihilation or decay?). Nuclear data on the \positron production XS at $\lesssim 5\%$ precision, particularly focussing on the $\sqrtsnn= 5\text{--}110$\,GeV energy range and transverse momenta $\pt < 1$\,GeV, should help discriminate among the different hypotheses. The search for photons is not hampered by the diffusion on magnetic fields, and regions expected to exhibit high DM densities can be targeted by telescopes, as pioneered by the observation from the Fermi-LAT experiment of a Galactic Centre Excess~\cite{Ajello_2016}, explainable as a DM manifestation. Measurements of the inclusive photon production, by exploiting collisions on H and He targets for $\sqrtsnn= 5\text{--}110$\,GeV, or at least its dominant contribution from $\pi^0$ decays, should be provided to correctly interpret Fermi-LAT observations.

\begin{figure}
    \centering
    \includegraphics[width=0.8\linewidth]{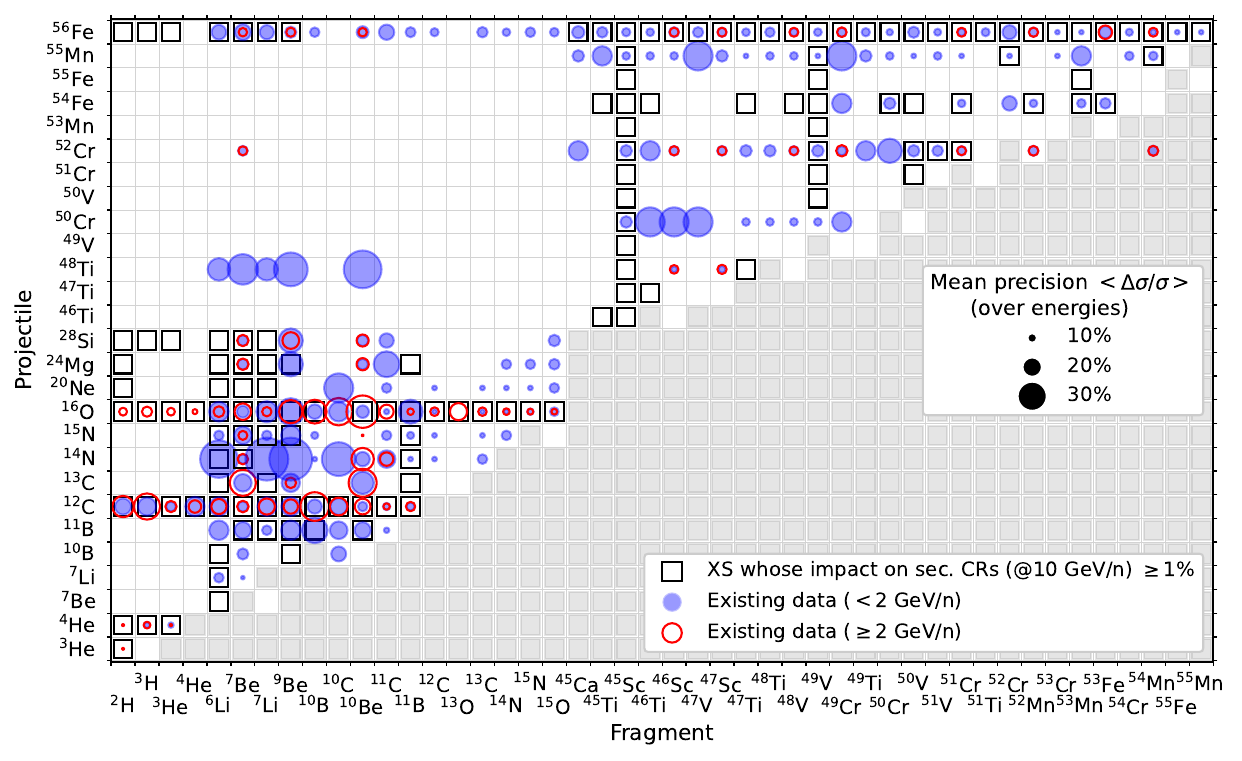}
    \caption{Existing nuclear data below (blue disks) and above 2\,GeV/n (red circles) and their relative precision (size of the circles), for reactions (black empty squares) contributing to at least 1\% of the flux of CR secondary species with $Z<30$.}
    \label{fig:fragmentation_XS}
\end{figure}

\paragraph{Production XS for secondary GCR species}
Uncertainties on the production XS of secondary GCR species -- namely \deuteron, $^3$He, LiBeB (and their isotopes), F, and sub-Fe elements, and also some GCR clocks (i.e., ratios of an unstable and a stable isotope of the same element, constraining GCR propagation time in the Galaxy) -- also impact the background calculations of all the above DM probes. In particular, the secondary \antiproton flux is impacted at the same level as that coming from the \antiproton production XS uncertainties \cite{Boudaud:2019efq}.
In addition, the astrophysical interpretation of the high-precision GCR data and its features (spectral breaks, anomalous abundance patterns, etc. -- for the fluxes of elements with atomic number $Z\leq30$, but also for the few measurements $Z>30$ --, is also limited by these production XS uncertainties. An overview of the available nuclear data for a H target, and more importantly on the missing XS, is illustrated in Fig.~\ref{fig:fragmentation_XS}. The required energy is from a few hundreds of MeV/n up to a few GeV/n, and ideally up to a few tens of GeV/n for a few reactions, in order to test the expected mild energy dependence of the XS. The precision needed is the percent level for the key reactions detailed in Ref.~\cite{XSCRC24_paper}. As shown in Ref.~\cite{2024PhRvC.109f4914G}, performing these measurements is a guaranteed game changer for the field.
Other than fragmentation XS, measurements of inelastic XS on both H and He targets are needed at the percent precision level (on H, less so on He) for all leading isotopes in GCR elements. Measurements for inelastic interactions on heavier materials (C, N, O, Al, Si, Fe, Cu) or crystal components (BGO and LYSO) are also needed at the $\lesssim 10\%$ precision, in order to solve the discrepant normalisation observed on several fluxes measured by AMS, CALET, DAMPE, etc. While CR experiments themselves have directly measured some of these XS~\cite{AMSXS,DAMPEXS}, the recent results by CERN experiments, as detailed in the following section, demonstrate how accelerator experiments can significantly contribute.

\section{A few successful examples of the HEP/astroparticle synergy}
\label{sec:synergies}
Given the very promising discovery potential described in the previous section, several HEP experiments at different facilities have started since a few years to provide the relevant XS measurements. In this section, a few examples of the successful synergy built between the astroparticle and the HEP communities are given, with a particular focus on CERN accelerators, together with some prospects for the future. It is worth to stress that many other nuclear facilities (existing or under construction), such as FAIR in Germany~\cite{FAIR}, Brookhaven in the USA~\cite{2016LSSR....9....2S}, and HIAF in China~\cite{Zhou:2022pxl}, could provide key measurements as well, in particular for societal topics requiring neutron, proton, He, O, Si, Fe, and even muons interactions on various targets (C, Al, Fe, etc.). These prospects are detailed in Ref.~\cite{XSCRC24_paper}. 

\subsection{CERN LHC experiments}
At CERN LHC accelerator, proton--proton (pp) collisions at energies between 7 and 13.6\,TeV are being provided to experiments, together with lead--lead or other ion species at a lower centre of mass energy. 

\paragraph{The LHCb experiment} 
The LHCb experiment is a single-arm spectrometer covering the pseudorapidity range $\eta \equiv -\ln[\tan(\theta/2)] \in [2, 5]$, being $\theta$ the angle with respect to the beam axis. It is composed of a tracking system giving a momentum resolution between 0.5 and up to 1\% at $p \sim 200$\,GeV/c, two Ring Cherenkov Imaging Detectors (RICH) providing charged hadron identification, a calorimeter and a muon system. Among all LHC experiments, LHCb is the only one also capable to operate in fixed-target configuration, by injecting gases in the accelerator through the System for Measuring Overlap with Gas (SMOG), notably H, D, and He, and studying beam-gas collisions with \mbox{$\sqrtsnn \in [27,113]$\,GeV}. As this energy matches the scales needed to interpret, e.g., AMS data, and the configuration reproduces that of a CR impinging on the ISM, LHCb equipped with SMOG can significantly contribute to the needs listed in Sec.~\ref{sec:motivation}.

\begin{figure}
    \centering
    \includegraphics[width=0.75\linewidth]{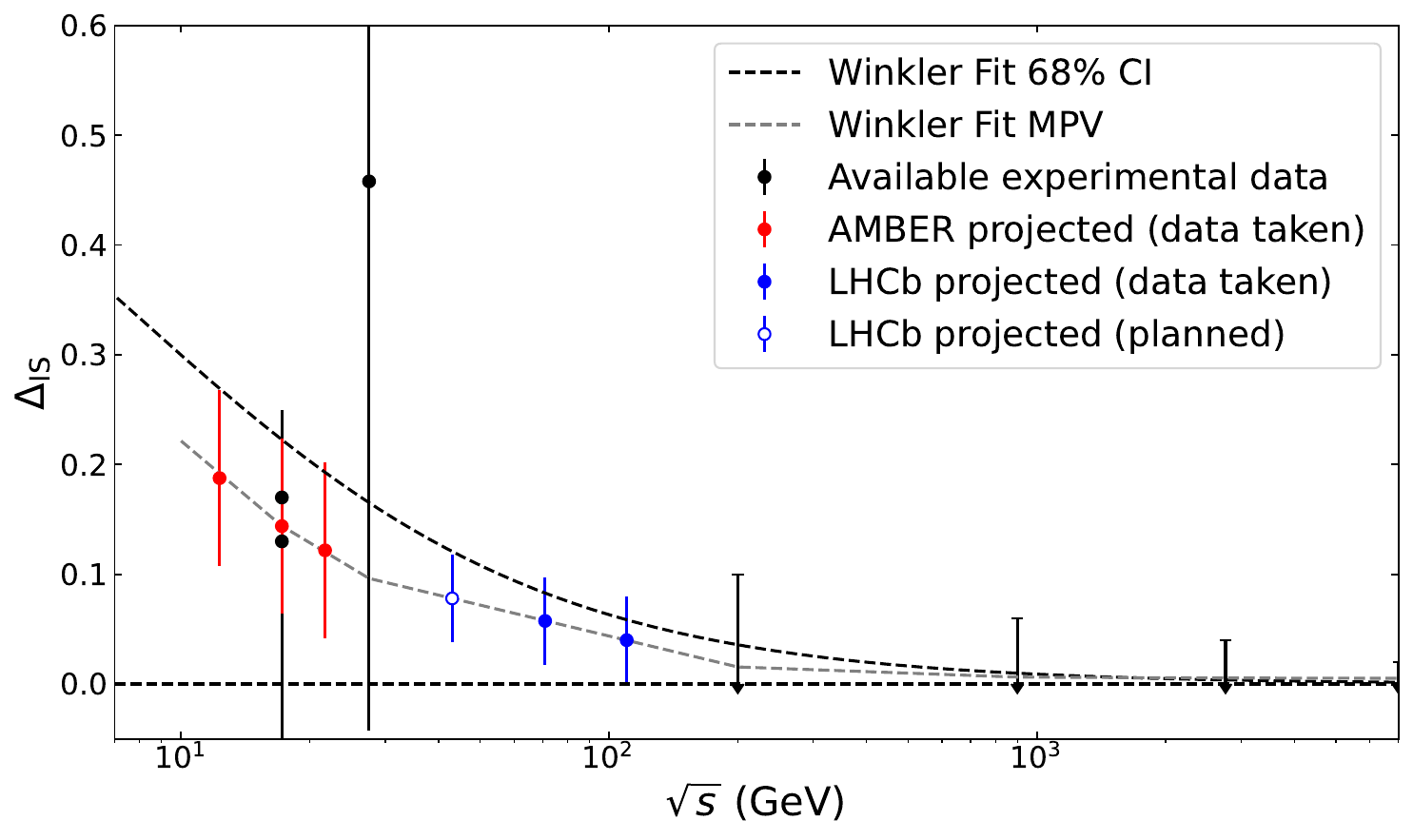}
    \caption{Projected impact of the AMBER (red) and LHCb (blue) measurements on the uncertainties of the isospin asymmetric production of \antiproton and \antineutron. Data points are filled in case of data already taken and empty in case of planned for the upcoming years. The black dashed line indicates the current uncertainty on $\Delta_{\rm IS}$, given the current nuclear data and model assumption (M.~Winkler~\cite{Winkler:2017xor}). The projected uncertainties for the individual measurement points from LHCb and AMBER are estimated based on assumed measurement uncertainties of $1\%$ and $2\%$ for the ratio of \antiproton production XSs in pD$_2$ and pH$_2$, respectively. Additionally, the limited phase-space coverage of both experiments reduces the sensitivity to the isospin asymmetry, which is primarily located in the target-fragmentation region (negative Feynman-x values, $x_{\rm F}$). As a conservative estimate, the given uncertainties account for sensitivities of the experiments only down to $x_{\rm F} = 0$. Other potential modelling uncertainties, such as an explicit $x_{\rm F}$ dependence of the isospin asymmetry, are not considered in this estimate.}
    \label{fig:antiproton_experiments}
\end{figure}

With proton--helium (pHe) data at $\sqrtsnn = 110$\,GeV collected in 2016, measurements of $\sigma_{\rm inv}^{\antiproton\;\rm prompt}$~\cite{LHCb-PAPER-2018-031} and $\Delta_{\Lambda}$~\cite{LHCb-PAPER-2022-006} have been published in 2018 and 2022, respectively. This first measurement ever for \antiproton production in pHe collisions allows discriminating between different parametrisations for the XS evolution as a function of \pt or Feynman-\textit{x}~\cite{Donato_NA61}. However, owing to the uncertainties on the pHe sample luminosity, lacking precise enough constraints on the injected gas pressure, an upgrade of the SMOG system, SMOG2~\cite{SMOG2_paper}, has been developed and installed in 2020. The luminosity uncertainty is projected to decrease from 6\% with SMOG to 1--2\% with the improved gas injection system offered by SMOG2. Since 2022, LHCb can continuously inject gas during data-taking, including non-noble species like H or D, and operate with two simultaneous interaction points at two different energy scales~\cite{SMOG2_paper}. During 2024, data samples with pHe, pH$_2$ and pD$_2$ collisions have been recorded at $\sqrtsnn = 70.9$\,GeV, and pHe and pH$_2$ data at $\sqrtsnn = 113$\,GeV as well. Very high-precision measurements for the \antiproton production XS in all these systems are foreseen. By comparing the \antiproton production in pH$_2$ and pD$_2$, in particular, the possible isospin excess between \antineutron and \antiproton production $\Delta_{\rm IS}$ will be constrained, as shown in Fig.~\ref{fig:antiproton_experiments}. Also, a request from LHCb for a short LHC data-taking period at lower collision energy, possibly down to 1\,TeV (corresponding to $\sqrtsnn = 43$\,GeV collisions in SMOG2), is being discussed. This would open the possibility to cover a larger fraction of the produced \antiproton spectrum, and constrain even more precisely the \antiproton XS listed in Table~\ref{tab:measurements}.

Light anti-nuclei identification was not initially foreseen at LHCb, but new methods have been developed to foster the discovery potential discussed in the previous section. The time-of-flight (TOF) capabilities of the 2015--2018 LHCb detector have been used to identify low-momentum (anti)d. Measurements of the (anti)d production XS, both absolute and relative to p/\antiproton, in $\sqrtsnn = 110$\,GeV pHe data, are ongoing. By exploiting the energy losses in several sub-detectors of the 2015--2018 LHCb detector, He identification in pp collisions has also been proven~\cite{HeliumID_LHCB}. Measurements of the He production XS in 13.6\,TeV and in $\sqrtsnn = 68.5$\,GeV fixed-target pNe data are ongoing. Moreover, the available (anti)He statistics in pp data has opened the possibility to measure in which processes these particles originate. While some prompt (anti)He has been found to originate in (anti)hypertriton decays~\cite{LHCb-CONF-2023-002}, limits for the (anti)He production in $\Lambda_b$ decays have been set~\cite{LHCb-CONF-2024-005}. These are particularly important for the topics discussed in Sec.~\ref{sec:motivation}, as they potentially exclude the proposed $\Lambda_b$ decay source of (anti)He candidates in GCRs \cite{Winkler_Lb}.

In the future, a further upgrade of the LHCb experiment is being discussed~\cite{LHCb-TDR-023}, to take data starting from 2036. Some scenarios for this include a dedicated TOF detector, named TORCH, which will greatly increase efficiencies for light nuclei identification.

\paragraph{The ALICE experiment} 
The ALICE experiment at LHC is a mid-rapidity experiment dedicated to nuclear physics in pp and heavy-ion collisions. The detector consists of several subdetectors that allow particle tracking in the pseudorapidity range of $|\eta| \leq 0.9$. Particle identification is mainly based on energy-loss measurements in a large time-projection chamber and is completed with a TOF measurement system for higher momenta.

Antiproton production has been measured in pp and PbPb collisions at several collision energies~\cite{2011_ALICE, ALICE2010, ALICE2020II}, although at higher scales with respect to those needed for CR data interpretation. Different antinuclei species and their ratios have also been measured, including \antideuteron, \antitriton, \antiheliumthree, and \antiheliumfour~\cite{Rasa2024, ALICE2022IV}. Additionally, (anti)hypernuclei -- nuclei including a nucleon with strangeness -- productions have been measured~\cite{Bartsch2023}.
The variety of measurements allows detailed studies of the formation process of light nuclei, often described by the coalescence model~\cite{PhysRevC.21.1301}. Parameters related to the coalescence probability of nuclei with two and three nucleons have been measured, and the experimental results have triggered studies beyond the classical coalescence model~\cite{ALICE:2021mfm, Mahlein2023}. Besides the formation process of antinuclei, ALICE also pioneered the measurement of \antideuteron and \antihelium absorption in matter~\cite{ALICE:dbar2020, ALICE2022II}. These processes have been experimentally mostly unexplored and impact the survival probability of antinuclei produced in our Galaxy during their propagation from their sources to the Earth.

ALICE has also performed momentum correlation studies~\cite{ALICE:2020ibs,ALICE:2023sjd,ALICE:2025aur}, known as femtoscopy technique, of several hadron pairs. This allowed to constrain the size of the particle-emitting source in pp or ion collisions, which is a necessary input for the coalescence model based on the Wigner function formalism. These results, together with the above-mentioned precise measurements of \antiproton and \antideuteron production spectra, showed that \antideuteron yields can be successfully predicted by this model~\cite{Mahlein2024}. Moreover, ALICE also used femtoscopy to study the residual strong interaction between hadron pairs and triplets, including hyperons. This provides essential inputs for astrophysics, for instance to constrain the equation of state of dense matter, and to understand better the composition of such dense systems, in particular the inner core of neutron stars~\cite{Fabbietti2020,ALICE:2022boj,Mihaylov:2023ahn}. Further details on this technique are given in Ref.~\cite{XSCRC24_paper}.

An upgrade of the experiment, ALICE3, is expected to start taking data by 2036. This will include a more extensive rapidity coverage, allowing to probe antinuclei production out of the central rapidity regime~\cite{ALICE2022III}.

\paragraph{The LHCf experiment}
The LHCf experiment \cite{LHCf:2008lfy} at LHC is made of two imaging and sampling calorimeters, covering a pseudorapidity region $\eta>8.4$. The experiment is dedicated to precise measurements of forward neutral particle production in pp and p--ion collisions, in order to provide calibration data that can be used to tune the hadronic interaction models used to simulate extensive air showers.

So far, the experiment has acquired data in pp collisions between $\sqrt{s} = 0.9$\ and $13.6$\,TeV and pPb collisions at $\sqrt{s_{\rm NN}} = 5.02$ and $8.16$\,TeV. The published results indicate a tension between models and data, which is particularly strong in the case of neutron production \cite{LHCf:2018gbv, LHCf:2020hjf}, but not negligible even in the case of $\gamma$ \cite{LHCf:2017fnw}, $\pi^0$ \cite{LHCf:2015rcj} and $\eta$ \cite{LHCf:2023yam} production. Thanks to an ongoing improvement of the reconstruction algorithm, it will be possible to measure $K^0_s$, and possibly $\Lambda^0$ forward production, from the data acquired in pp collisions at $\sqrt{s} = 13.6$\,TeV. In parallel, the LHCf--ATLAS joint analysis will give a better insight in the understanding of production mechanisms, leading to greater constraints for the calibration of hadronic interaction models. The LHCf measurements can be used to constrain production XSs entering the calculation of the astrophysical $\gamma$-ray background (from GCRs on the ISM), as shown in Ref.~\cite{orusa2023new}. Indeed, $\pi^0$ and $\eta$ are the main contributions to the $\gamma$-ray flux, and the inclusive $\gamma$ production measured by the experiment can be used as a benchmark.

\subsection{CERN SPS and PS experiments}
Before circulating in LHC, particles are accelerated at CERN by lower-energy machines. The Proton-Synchrotron (PS) accelerator is one of the first acceleration stages, and can reach a maximum energy of 26\,GeV. It delivers particles to the Super Proton-Synchrotron (SPS), which has a maximum energy of 450\,GeV. Secondary beams of nuclei can also be provided to experiments, from 10 to 158\,GeV energy per nucleon. To this purpose, a primary beam of $^{208}$Pb is extracted onto a beryllium target and nuclear fragments are guided to the experimental area. The rigidity acceptance of the beam line can be adjusted to select the specific mass-to-charge ratio of the desired nuclei. The SPS also delivers secondary hadrons at momenta up to 400\,GeV/c, depending on the beam line\footnote{\url{https://sba.web.cern.ch/sba/BeamsAndAreas/H2/H2_presentation.html}}: on the H2 beam line (for NA61/SHINE), secondary hadron beams up to 400\,GeV/c can be produced; on the M2 beam line (for AMBER), 280\,GeV/c is the maximum hadron beam momentum.

\paragraph{The AMBER experiment}
The AMBER experiment at the M2 secondary beam line of CERN SPS is a fixed-target experiment that started data-taking in 2023, as the successor of the long-standing COMPASS experiment. Within the first approved phase of the experiment from 2023 to around 2031, AMBER reused and upgraded the 2-stage magnetic spectrometer from COMPASS. The experimental setup includes two differential Cherenkov counters, with achromatic ring focus to identify protons in the mixed hadron beam, a cryogenic target filled with the target gas, and the AMBER spectrometer to characterise the particles created in the interaction.
In order to measure their momentum, the AMBER spectrometer consists of around 300 tracking detector planes and two spectrometer magnets, with a bending strength of up to $1$\,Tm and 4\,Tm, respectively. Additionally, a RICH detector and muon detectors allow particle identification over an extensive momentum range. Antiprotons with a total momentum between $10$\,GeV/c and $60$\,GeV/c, and transverse momentum up to $2$\,GeV/c, are identified.
In 2023, a cryogenic target filled with He was used to acquire data aiming to a \antiproton production measurement in pHe collisions. Data were recorded at six different collision energies between $\sqrtsnn = 10.7$\,GeV and $\sqrtsnn = 21.7$\,GeV. In 2024, a new cryogenic target was built to allow the usage of flammable gases, such as H and D. For both targets, collisions at $12.3$\,GeV, $17.3$\,GeV, and $21.7$\,GeV were recorded, with an identical spectrometer setup. Besides providing \antiproton production XS for the different targets with about $5\%$ relative uncertainty, one dedicated goal of the measurements is to investigate the aforementioned possible isospin asymmetry of \antiproton to \antineutron production, comparing the \antiproton production XS in pH$_2$ and pD$_2$ collisions. The expected uncertainties on the individual XS should allow a measurement of the isospin asymmetry $\Delta_{IS}$, at the $10\%$ level for the three collision energies, as illustrated in Fig.~\ref{fig:antiproton_experiments}. In the case of a measurable asymmetry, the measurement of the different collision energies and the combination with LHCb data would additionally constrain the collision-energy dependence of the effect.

In the future, the AMBER spectrometer will undergo several upgrades and improvements to operate the spectrometer at around 10--100 times higher read-out rates~\cite{Zemko2021}. This improvement would allow the measurement of rare particles, such as \antideuteron and \antihelium. However, dedicated nuclei identification is needed for these studies and is currently under investigation.

\paragraph{The NA61/SHINE experiment}
The fixed-target experiment NA61/SHINE at the CERN SPS is a hadron spectrometer capable of studying collisions of hadrons with different targets, over a wide range of incident beam momenta~\cite{na61}. It is the successor to the NA49 experiment, which pioneered \antiproton production XS measurements at $\sqrtsnn = 17.3$\,GeV in pp and pC collisions, covering nearly the full phase space of created \antiproton, and which provided a first measurement of the potential isospin asymmetric production of \antiproton and \antineutron.

NA61/SHINE consists of different subdetectors for particle identification. It already recorded pp interactions with beam momenta from 13 to 400\,GeV/c, and also collected data for other hadron interactions, including pC, $\pi^\pm$C, ArSc, pPb, BeBe, XeLa, PbPb at different energies. Between 2018 and 2022, upgrades to the time projection chamber backend electronics resulted in improvements in the specific energy loss resolution. Essential for future \antideuteron production measurements is the new data acquisition system, with about 20~times faster rate, and new TOF detectors with improved time resolution~\cite{Gazdzicki:2692088, Fields:2739340}. NA49 and NA61/SHINE have published several relevant
data~\cite{NA49:2012jna, Aduszkiewicz:2017sei} to tune CR antinuclei formation, as well as data
relevant for the tuning of models describing CR-induced air showers~\cite{NA61SHINE:2017vqs, NA61SHINE:2022tiz}.

With respect to \antiproton production and antinuclei formation, the published measurements of light nuclei in pp and various nucleus--nucleus data sets can be used to study the production of light ions at the threshold. These measurements will complement the NA49~\cite{dbarna49,Anticic:2016ckv} and ALICE results and allow testing coalescence and thermal models in a different regime.
Extended data-taking with an upgraded NA61/SHINE experiment relevant to understanding cosmic antinuclei is already planned before 2026. A pp dataset of approximately 600~M events, collected with a beam energy of 300\,GeV, will provide new measurements of pp correlations, \antiproton, and deuterons. This proposed dataset will feature significantly reduced systematic and statistical uncertainties, enhancing the ability to discriminate between different nuclear formation models. It is also anticipated that, for the first time, \antideuteron will be identified in this range crucial for the cosmic \antideuteron interpretation. Combining these new measurements will enable building, testing, and validating data-driven \deuteron and \antideuteron production models in the energy range most relevant to CRs data interpretation. 

NA61/SHINE is also pursuing a rich programme of nuclei fragmentation measurements. A first pilot run of carbon fragmentation measurement at 13.5\,$A$\,GeV was conducted in 2018, demonstrating that the measurements are possible~\cite{NA61SHINE:2024rzv}. For this type of measurement, the primary $^{208}$Pb is extracted from the SPS and fragmented in collisions with a 160\,mm-long Be plate in the H2 beam line. The resulting nuclear fragments of a chosen rigidity are guided to the NA61/SHINE experiment, where the projectile isotopes are identified, via a measurement of the particle charge and TOF over a length of approximately $240$\,m. Data of the fragmentation of nuclei from Li to Si at 12.5\,GeV/n were collected at the end of 2024 and are currently being analysed. NA61/SHINE also provides critical data for inelastic XSs\cite{NA61SHINE:2019aip}, which need to be measured at a few percent precision, see Ref.~\cite{XSCRC24_paper}.
In the future, these measurements of nuclear fragmentation with NA61/SHINE can potentially be extended up to Fe, and be performed at different energies.

\paragraph{The n\_\!TOF experiment}
The n\_\!TOF neutron facility~\cite{ntof} relies on PS proton beam collisions on a thick lead target to create a neutron flux ranging from the thermal region to several GeV. The facility was optimised for high-precision measurements of radioactive materials, leading to a very low duty cycle and very long flight-paths, from 20 to 180\,m.
In recent years, developments have been made to measure XS of reactions leading to the emission of charged particles using silicon detectors. Up to now, only preliminary results were obtained, up to a few MeV. The main limitation is the very strong $\gamma$ flash that comes with every neutron pulse and blinds most detectors for a short time. A development has started to use gaseous detectors for such measurements, since these detectors are much less sensitive to the $\gamma$ flash.
Measurements on various targets (C, O, etc.) could be crucial for cosmogenic studies, where neutron-induced reactions are the dominant contributors to the formation, for instance, of $^{10}$Be, see Ref.~\cite{XSCRC24_paper}.


\section{Conclusions and roadmap for the future}
\label{sec:conclusions}
Owing to the percent-level precision era CR physics is currently living, \textbf{potential for impactful and even ground-changing discoveries is mostly limited by the precision at which hadron and nuclear XSs are known}. 
This is going to be a long-lasting and growing issue, owing to the rich CR programme and projects for the next decades, with the AMS upgrade, DAMPE, and CALET taking data until 2030, the ongoing HELIX~\cite{HELIX:Coutu_2024} and GAPS~\cite{Aramaki_2016} balloon flights (for cosmic-ray clocks and anti-matter detection, respectively), the TIGER-ISS~\cite{2024icrc.confE.171R} and HERD~\cite{Kyratzis:2022rxa} detectors planned to run in 2027--2028, and the more uncertain ALADInO~\cite{2021ExA....51.1299B,ALADInO:2022ntw} and AMS-100~\cite{Schael:2019lvx} projects beyond 2040.

Particle physics experiments, mostly at CERN facilities, have started since a few years to address this issue, building very successful synergies with the astrophysics community exemplified by initiatives like the XSCRC workshop series. Much more is needed, though: in order to establish a DM evidence from CR space-borne experiments data, antimatter production XSs for proton collisions on H or He targets must be provided with uncertainties below 5\%, together with production XSs of LiBeB (or other GCR secondary nuclei) from C, O, Si, Fe with uncertainties below a few percents. Inelastic XSs at a few percent precision are also needed for interpreting GCR data and accounting for interactions in CR detectors. In many cases, such XSs, or complementary ones, also enable crucial societal topics, such as cosmogenic production in meteorites, space exploration and radiation protection, and hadrontherapy, to progress in solving their associated puzzles.

The main message this contribution wants to convey is that support by the community, and surely in the 2026 European Strategy for Particle Physics Update, is vital to keep harvesting on the discussed success examples. On the modelling side, a robust and comprehensive compilation of missing and needed high-precision nuclear XS data, for CR physics but also for other applications, have been provided in Ref.~\cite{XSCRC24_paper}.
On the nuclear data side, the possibilities offered by current facilities and experiments are being surveyed more closely, and new opportunities and detectors must be prepared and planned for to make strong proposals for beam time dedicated to these XS measurements. Given the wide range of energies, hundreds of MeV to hundreds of TeV, projectiles (all nuclei, and possibly neutrons and anti-matter), targets (mostly H and He), and fragments to measure, not a single facility will provide all the needs, but all new inputs might have a significant impact in advancing in the field. We hope this paper will give more visibility to the current efforts and motivate experimentalists to join, in order to perform at least the most urgent measurements in our wish lists. 

The clear, long term, challenging but rewarding, programme ahead of us, indeed, faces the difficulties of limited human resources and funding, in a future constrained by a necessary decrease of our footprint in terms of greenhouse gas emissions. The advantage of our programme is that it mostly relies on existing facilities, taking a priori a very small fraction of the physics programmes for which the latter were conceived. In this respect, given the broad and interdisciplinary questions, we hope that the road map provided in this paper will help convince and gather support from deciding committees and the many agencies financing research world-wide.

\section*{Acknowledgements}
We thank our colleagues of the XSCRC series for lively discussions over the years, which helped to crystallise and define the XS needs for various physics cases, leading to new XSs measurements, and many successful synergies between our communities. We thank CERN, and in particular the Theoretical Physics Department, for hosting and supporting all the editions of the XSCRC workshop.
We thank R.~Delorme and V.~Tatischeff for their helpful suggestions regarding nuclear and medical physics experts for the XSCRC\,2024 workshop.

F.~Donato is supported by the 
Research grant {\sl The Dark Universe: A Synergic Multimessenger Approach}, No.~2017X7X85K, funded by the {\sc Miur}.
P.~Coppin and A.~Tykhonov acknowledge the support of the European Research Council (ERC) under the European Union’s Horizon~2020 research and innovation programme (Grant No.~851103). P. Coppin is also supported by the Swiss National Science Foundation (SNSF).
Philip von Doetinchem knowledges the support of the National Science Foundation grant~PHY-2411633.
F.~Donato, C.~Evoli and M.~Di~Mauro acknowledge support from the research grant {\sl TAsP (Theoretical Astroparticle Physics)} funded by Istituto Nazionale di Fisica Nucleare (INFN).
Y.~Génolini et P.~D.~Serpico acknowledge support by the Université Savoie Mont Blanc via the research grant NoBaRaCo.
P. Ghosh was supported by the Pioneers mission, the Trans-Iron Galactic Element Recorder for the International Space Station, TIGERISS, an Exceptional Nucleosynthesis Pioneer.
The work of D.~M.~Gomez~Coral is supported by UNAM-PAPIIT IA101624 and SECIHTI under grant CBF2023-2024-118.
M.~Mahlein is supported by the European Research Council (ERC) under the European Union’s Horizon 2020 research and innovation programme (Grant Agreement No~950692) and the BMBF 05P24W04 ALICE.
The work of I.~Leya is funded by the Swiss National Science Foundation (200020\_219357, 200020\_196955).
The work of M.J.~Losekamm is funded by the Deutsche Forschungsgemeinschaft (DFG, German Research Foundation) via Germany's Excellence Initiative -- EXC-2094 -- 390783311.
The work of D.~Maurin was supported by the {\em Action thématique Phénomènes extrêmes et Multi-messagers (AT-PEM)} of CNRS/INSU PN Astro with INP and IN2P3, co-funded by CEA and CNES, and by the INTERCOS project funded by IN2P3.
J.W.~Norbury was supported by the RadWorks project within Exploration Capabilities of the Mars Campaign Development Division in the Exploration Systems Development Mission Directorate of NASA, United States.
J.~Ocampo-Peleteiro and A.~Oliva are supported by INFN and ASI under ASI-INFN Agreements No.~2019-19-HH.0, No.~2021-43-HH.0, and their amendments.
L.~Orusa acknowledges the support of the Multimessenger Plasma Physics Center (MPPC), NSF grants~PHY2206607.
M.~Unger acknowledges the support of the German Research Foundation DFG (Project No.~426579465).
M.-J.~Zhao acknowledges the support from the National Natural Science Foundation of China under Grants No.~12175248 and No.~12342502.

\bibliographystyle{elsarticle-num-names}
\bibliography{main}
\end{document}